# Symmetric Boolean Function with Maximum Algebraic Immunity on Odd Number of Variables[*]

Na Li, Wen-feng Qi

Department of Applied Mathematics, Zhengzhou Information Engineering University

P.O.Box 1001-745, Zhengzhou, 450002, People's Republic of China

E-mail: mylina_1980@yahoo.com.cn, wenfeng.qi@263.net

**Abstract.** To resist algebraic attack, a Boolean function should possess good algebraic immunity (*AI*). Several papers constructed symmetric functions with the maximum algebraic immunity $\left\lceil \frac{n}{2} \right\rceil$. In this correspondence we prove that for each odd *n*, there is exactly one trivial balanced *n*-variable symmetric Boolean function achieving the algebraic immunity $\left\lceil \frac{n}{2} \right\rceil$. And we also obtain a necessary condition for the algebraic normal form of a symmetric Boolean function with maximum algebraic immunity.

## 1. Introduction

The set of symmetric Boolean functions is an interesting subset of Boolean functions which have the function values determined by the weight of the vector. Symmetric functions can be represented in a very compact way both for their algebraic normal forms and for their value vectors. This property considerably reduces the amount of memory required for storing the function and is of great interest in software applications. Also in hardware implementation, only a low number of gates are required[1]. Many classical cryptographic properties of symmetric

[*] This work was supported by the National Natural Science Foundation of China (Grant 60373092).

Boolean functions were studied. For example, maximum nonlinearity of symmetric Boolean functions on odd number of variables has been studied in [2], properties such as resiliency, propagation criterion of symmetric Boolean functions have been studied in [3].

Recently algebraic attack has become an important tool in cryptanalysing stream and block cipher systems. The attack recovers the secret key by solving overdefined systems of multivariate equations[4-8]. A new cryptographic property for designing Boolean functions to resist this kind of attack called algebraic immunity[9-13] was proposed. It has been proved in [5, 9] that the algebraic immunity of an $n$-variable Boolean function is upper bounded by $\left\lceil \frac{n}{2} \right\rceil$.

It is important and significant to search or construct Boolean function with maximum algebraic immunity. Several papers[12, 13] constructed symmetric functions achieving maximum algebraic immunity. A symmetric Boolean function on odd number of variables with maximum algebraic immunity has been constructed in [12,13]. The problem is whether there are any other Symmetric Boolean functions on odd number of variables that possess this property. In this correspondence, we prove that for each odd $n$, there is exactly one trivial balanced $n$-variable symmetric Boolean function achieving the algebraic immunity $\left\lceil \frac{n}{2} \right\rceil$. In addition, we obtain a necessary condition for the algebraic normal form of an $n$-variable symmetric Boolean function with maximum algebraic immunity for any positive integer $n$.

## 2. Preliminaries

Let $\mathbf{F}_2^n$ be the set of all $n$-tuple of elements in the finite field $\mathbf{F}_2$. A Boolean function of $n$ variables is a mapping from $\mathbf{F}_2^n$ into $\mathbf{F}_2$. The symbol "+" and "$\sum$" in this correspondence are referred to the sum over $\mathbf{F}_2$.

An $n$-variable Boolean function $f$ can be uniquely represented by the truth table which is the vector of length $2^n$ consisting of its function values. $f$ is balanced if the truth table contains an equal number of 1's and 0's. Let $S = (s_1, s_2, \ldots, s_n) \in \mathbf{F}_2^n$, the Hamming weight of $S$ is the number of 1's in $\{s_1, s_2, \ldots, s_n\}$.

Another unique representation of an $n$-variable Boolean function $f$, called the algebraic normal form (ANF), is a polynomial in $\mathbf{F}_2[x_1, \ldots, x_n]/(x_1^2 - x_1, \ldots, x_n^2 - x_n)$.

$$f(x_1, \ldots, x_n) = a_0 + \sum_{1 \leq i \leq n} a_i x_i + \sum_{1 \leq i < j \leq n} a_{i,j} x_i x_j + \ldots + a_{1,2,\ldots n} x_1 x_2 \ldots x_n$$

where the coefficients $a_0, a_i, a_{i,j}, \ldots, a_{1,2,\ldots n} \in \mathbf{F}_2$. The algebraic degree denoted by $\deg(f)$, is the number of variables in the highest order term with non zero coefficient.

Now, let we introduce some definitions and basic properties on symmetric Boolean functions.

**Definition 1**[3]. An $n$-variable Boolean function $f$ is to be symmetric if its output is invariant under any permutation of its inputs bits, i.e.

$$f(x_1, \ldots, x_n) = f(x_{\tau(1)}, \ldots, x_{\tau(n)})$$

for all permutations $\tau$ of $\{1, \ldots, n\}$.

This means that symmetric Boolean function $f$ has the property that the function value of all vectors with the same weight is equal. As a consequence, the truth table of an $n$-variable symmetric function can be replaced by a vector $v_f = (v_f(0), \ldots, v_f(n)) \in \mathbf{F}_2^{n+1}$ where the components $v_f(i)$ represents the function value for vectors of weight $i$ for $0 \leq i \leq n$. The vector $v_f$ is called simplified value vector of $f$.

Also the ANF representation for a symmetric function can be replaced by a shorter form[3], called simplified ANF (SANF). Denote the $n$-variable homogeneous symmetric Boolean function, which contains all terms of degree $i$ for $0 \leq i \leq n$, by $\sigma_i^{(n)}$. Then the SANF is a polynomial in $\mathbf{F}_2[x_1, \ldots, x_n]/(x_1^2 - x_1, \ldots, x_n^2 - x_n)$ can be written as follows:

$$f(x_1, \ldots, x_n) = \sum_{i=0}^{n} \lambda_f(i) \sigma_i^{(n)}, \quad \lambda_f(i) \in \mathbf{F}_2.$$

The vector $\lambda_f = (\lambda_f(0), \ldots, \lambda_f(n)) \in \mathbf{F}_2^{n+1}$ is called the simplified ANF vector (SANF vector).

Let $a$ and $b$ be two positive integers with their 2-adic representations $a = \sum_{l=0}^{m} a_l 2^l$

and $b = \sum_{l=0}^{m} b_l 2^l$. $a \preceq b$ means that for any $l$, $a_l \leq b_l$. The relationship between the simplified value vector and the simplified ANF vector can be obtained.

**Lemma 1**[3]. Let $f$ be an $n$-variable symmetric Boolean function. Then, its simplified value vector $v_f$ and its simplified ANF vector $\lambda_f$ are related by

$$v_f(i) = \sum_{k \preceq i} \lambda_f(k) \text{ and } \lambda_f(i) = \sum_{k \preceq i} v_f(k) \text{ for any } i \in \{0, 1, \ldots, n\}.$$

A large number of balanced symmetric Boolean functions have particular property of their simplified value vector.

**Definition 2**[3]. Let $n$ be odd and $f$ be an $n$-variable symmetric Boolean function. $f$ is called a trivial balanced function if

$$v_f(i) = v_f(n - i) + 1 \text{ for any } 0 \leq i \leq n$$

The exhaustive search for all balanced symmetric functions up to 128 variables presented in [14] shows that, for odd $n$, all balanced symmetric functions are trivial balanced except for $n \in \{13, 29, 31, 33, 35, 41, 47, 61, 63, 73, 97, 103\}$. Even for $n$ belonging to the last set, the number of non trivial balanced symmetric Boolean functions is expected small.

## 3. Algebraic Immunity and Previous Work

Now let us focus on the cryptanalysis property of Boolean functions to resist algebraic attack. A Boolean function should be of high algebraic degree to be cryptographically secure. Further more, to resist algebraic attack, the function should not have a low degree multiple[5]. Now it is clear that it is sufficient to consider the annihilator of both $f$ and $f + 1$[9].

**Definition 3**[9]. For a given $n$-variable Boolean function $f$, a nonzero $n$-variable Boolean function $g$ is called an annihilator of $f$ if $f \cdot g = 0$.

**Definition 4**[9]. For a given $n$-variable Boolean function $f$, the algebraic immunity $AI(f)$ is the minimum value of $d$ such that $f$ or $f + 1$ admits an annihilating function of degree $d$.

Actually, the $AI$ is upper bounded by $\left\lceil \dfrac{n}{2} \right\rceil$ which has been shown in [5,9]. There is an increasing interest in construction of Boolean functions with good algebraic immunity[11-13]. Symmetric Boolean functions with maximum algebraic immunity were constructed by different methods in [12,13].

**Lemma 2**[10]**.** Let $n$ be odd and $f$ be an $n$-variable Boolean function, $AI(f) = \left\lceil \dfrac{n}{2} \right\rceil$ implies that $f$ is balanced.

**Lemma 3**[12,13]**.** Let $n$ be odd and $f$ be an $n$-variable symmetric Boolean function, if its simplified value vector $v_f$ satisfies

$$v_f(i) = \begin{cases} a & \text{for } i \leq \left\lfloor \dfrac{n}{2} \right\rfloor \\ a+1 & \text{for } i > \left\lfloor \dfrac{n}{2} \right\rfloor \end{cases}$$

where $a \in \mathbf{F}_2$. Then $AI(f) = \left\lceil \dfrac{n}{2} \right\rceil$.

It is clear that $AI(f) = AI(f+1)$, so we do not distinguish $f$ and $f+1$ with regard to algebraic immunity. For even $n$, several classes of $n$-variable symmetric Boolean functions with maximum algebraic immunity were constructed. For odd $n$, only one such symmetric Boolean function mentioned in Lemma 3 was constructed. The problem is whether there are any other symmetric Boolean functions on odd number of variables with maximum algebraic immunity. In the next section we will prove that there is exactly one symmetric Boolean function with maximum algebraic immunity among trivial balanced symmetric Boolean functions on odd number of variables.

## 4 Main Result

At first, we construct a class of symmetric Boolean functions whose value is equal to 1 only possible for vectors of certain weights.

**Lemma 4.** For any positive integer $i$, there is at least one nonzero $(2i + 1)$-variable symmetric Boolean function $g$ which satisfies

(i) $\deg(g) \leq i$.

(ii) $v_g(0) = v_g(i + 1) = \ldots = v_g(i + i) = 0$.

*Proof:* The requirement of algebraic degree of $g$ implies that

$$\lambda_g(i + 1) = \ldots = \lambda_g(2i + 1) = 0. \tag{1}$$

So we can assume $g = \lambda_g(0) + \lambda_g(1)\sigma_1^{(2i+1)} + \ldots + \lambda_g(i)\sigma_i^{(2i+1)}$.

Next we will show that there is at least one nonzero such $g$ whose simplified value vector satisfies

$$v_g(0) = v_g(i + 1) = \ldots = v_g(i + i) = 0. \tag{2}$$

Let $i = 2^t + j$, where $t \geq 0$, $0 \leq j < 2^t$. Then by lemma 1, we have

$$v_g(i + (2^t - j)) = v_g(2^{t+1}) = \sum_{k \leq 2^{t+1}} \lambda_g(k) = \lambda_g(0) = v_g(0). \tag{3}$$

By lemma 1 and (1), each one of $v_g(0)$, $v_g(i + 1)$, ..., $v_g(i + i)$ can be represented as a linear combination of $\lambda_g(0)$, $\lambda_g(1)$, ..., $\lambda_g(i)$. We consider $\lambda_g(0)$, $\lambda_g(1)$, ..., $\lambda_g(i)$ as $(i + 1)$ variables. By (2) and (3) we can obtain $i$ linear equations on these variables. So there must exist at least one nonzero solution. □

**Remark.** Actually, by the proof of lemma 4, we can construct nonzero symmetric Boolean functions satisfying the requirement by solving systems of linear equations. Note that the value of such $(2i + 1)$-variable symmetric Boolean functions is equal to 1 only possible for the vectors of weight 1, 2, ..., $i$ and $2i + 1$.

**Example.** We obtain the following symmetric Boolean functions by solving the systems of linear equations.

For $i = 1$, we get $g_1(x_1, x_2, x_3) = \sigma_1^{(3)} = x_1 + x_2 + x_3$. $v_{g_1} = (0, 1, 0, 1)$, and $\deg(g_1) = 1$.

For $i = 2$, we get $g_2(x_1, x_2, x_3, x_4, x_5) = \sigma_1^{(5)} + \sigma_2^{(5)} = x_1 + \ldots + x_5 + x_1x_2 + \ldots + x_4x_5$. $v_{g_2} = (0, 1, 1, 0, 0, 1)$, and $\deg(g_2) = 2$.

For $i = 3$, we get $g_3(x_1, x_2, x_3, x_4, x_5, x_6, x_7) = \sigma_3^{(7)} = x_1x_2x_3 + \ldots + x_5x_6x_7$. $v_{g_3} = (0, 0, 0, 1, 0, 0, 0, 1)$, and $\deg(g_3) = 3$.

For $i = 4$, we get $g_4(x_1, x_2, x_3, x_4, x_5, x_6, x_7, x_8, x_9) = \sigma_1^{(9)} + \sigma_2^{(9)} + \sigma_3^{(9)} + \sigma_4^{(9)} = x_1 + \ldots + x_9 + \ldots + x_1x_2x_3x_4 + \ldots + x_6x_7x_8x_9$. $v_{g_4} = (0, 1, 1, 1, 1, 0, 0, 0, 0, 1)$, and $\deg(g_4)$

= 4.

For $i = 5$, we get $g_5(x_1, x_2, x_3, x_4, x_5, x_6, x_7, x_8, x_9, x_{10}, x_{11}) = \sigma_3^{(11)} + \sigma_5^{(11)}$. $v_{g_5} = (0, 0, 0, 1, 0, 1, 0, 0, 0, 0, 1)$, and $\deg(g_5) = 5$.

We will give the main result in the next two theorems.

**Theorem 1.** Let odd $n \geq 1$ and $f$ be a trivial balanced $n$-variable symmetric Boolean function. If $AI(f) = \left\lceil \frac{n}{2} \right\rceil$, then

$$v_f(0) = \ldots = v_f(\left\lfloor \frac{n}{2} \right\rfloor) = v_f(\left\lceil \frac{n}{2} \right\rceil) + 1 = \ldots = v_f(n) + 1.$$

*Proof:* Because $f$ is a trivial balanced symmetric Boolean function, it is sufficient to show that $v_f(0) = \ldots = v_f(\left\lfloor \frac{n}{2} \right\rfloor)$.

Next we will show $v_f(\left\lfloor \frac{n}{2} \right\rfloor - k + 1) = \ldots = v_f(\left\lfloor \frac{n}{2} \right\rfloor)$ holds for all integers $1 \leq k \leq \left\lfloor \frac{n}{2} \right\rfloor + 1$.

For $k = 1$, it is apparent that $v_f(\left\lfloor \frac{n}{2} \right\rfloor - k + 1) = v_f(\left\lfloor \frac{n}{2} \right\rfloor)$. Now suppose $i$ ($1 \leq i \leq \left\lfloor \frac{n}{2} \right\rfloor + 1$) is the largest integer such that $v_f(\left\lfloor \frac{n}{2} \right\rfloor - i + 1) = \ldots = v_f(\left\lfloor \frac{n}{2} \right\rfloor)$. If $i = \left\lfloor \frac{n}{2} \right\rfloor + 1$, then $v_f(0) = \ldots = v_f(\left\lfloor \frac{n}{2} \right\rfloor)$, and Theorem 1 holds. In the following, we only consider the case when $1 \leq i \leq \left\lfloor \frac{n}{2} \right\rfloor$.

By the definition of $i$, we know $v_f(\left\lfloor \frac{n}{2} \right\rfloor - i + 1) \neq v_f(\left\lfloor \frac{n}{2} \right\rfloor - i)$, so we have

$$v_f(\left\lfloor \frac{n}{2} \right\rfloor - i + 1) = 1 + v_f(\left\lfloor \frac{n}{2} \right\rfloor - i) = v_f(\left\lfloor \frac{n}{2} \right\rfloor + i + 1).$$

The second "=" holds because $f$ is trivial balanced symmetric function.

Thus

$$v_f(\lfloor \frac{n}{2} \rfloor - i + 1) = \ldots = v_f(\lfloor \frac{n}{2} \rfloor) = v_f(\lfloor \frac{n}{2} \rfloor + i + 1). \qquad (4)$$

By lemma 4, we can construct a nonzero symmetric Boolean function $g$ on variables $x_1, x_2, \ldots, x_{2i+1}$, such that $g(x_1, x_2, \ldots, x_{2i+1})$ equal one only possible for vectors in $\mathbf{F}_2^{2i+1}$ of weight $1, 2, \ldots, i$ and $2i + 1$, and the degree of $g$ is at most $i$. Let $h(x_{2i+2}, \ldots, x_n) = (x_{2i+2} + x_{2i+3})\cdots(x_{n-1} + x_n)$. It is apparent that the value of $h$ is one only for a subset of vectors in $\mathbf{F}_2^{n-2i-1}$ of weight $(n - 2i - 1)/2 = \lfloor \frac{n}{2} \rfloor - i$. And the degree of $h$ equals $(n - 2i - 1)/2 = \lfloor \frac{n}{2} \rfloor - i$. Consequently the value of $n$-variable Boolean function $g \cdot h$ is one only possible for vectors in $\mathbf{F}_2^n$ of weight $\lfloor \frac{n}{2} \rfloor - i + 1, \ldots, \lfloor \frac{n}{2} \rfloor$ and $\lfloor \frac{n}{2} \rfloor + i + 1$.

By (4) we know, $v_f(\lfloor \frac{n}{2} \rfloor - i + 1) = \ldots = v_f(\lfloor \frac{n}{2} \rfloor) = v_f(\lfloor \frac{n}{2} \rfloor + i + 1) = 0$, or $1$. For the former, $g \cdot h$ is the annihilator of $f$, and for the latter, $g \cdot h$ is the annihilator of $f + 1$. And the algebraic degree of $g \cdot h$ is at most $\lfloor \frac{n}{2} \rfloor - i + i = \lfloor \frac{n}{2} \rfloor$, which contradicts $AI(f) = \lceil \frac{n}{2} \rceil$. □

**Theorem 2.** Let $n$ be odd and $f$ be an $n$-variable symmetric Boolean function which is trivial balanced. Then $AI(f) = \lceil \frac{n}{2} \rceil$ if and only if $v_f$ satisfies

$$v_f(i) = \begin{cases} a & \text{for } i \leq \lfloor \frac{n}{2} \rfloor \\ a+1 & \text{for } i > \lfloor \frac{n}{2} \rfloor \end{cases}$$

where $a \in \mathbf{F}_2$.

*Proof:* It is obvious by theorem 1 and lemma 3. □

In addition, we obtain a necessary condition for the SANF of a symmetric Boolean function with maximum algebraic immunity.

**Theorem 3**. Let integer $2^t \leq n < 2^{t+1}$, $t \geq 0$, and $f$ be an $n$-variable symmetric Boolean function. If $AI(f) = \left\lceil \dfrac{n}{2} \right\rceil$, then at least one of $\lambda_f(2^{t-1})$, $\lambda_f(2^t)$ and $\lambda_f(2^t + 2^{t-1})$ is 1.

*Proof:* Let $0 < a \leq n$, $a = 2^{a_1} + \ldots + 2^{a_l}$ is the 2-adic representation of $a$, where $0 \leq a_1 < \ldots < a_l \leq t$, $1 \leq l \leq t$. Then for any $0 \leq b \leq n$

$$a \preceq b \iff 2^{a_1} \preceq b, \ldots, 2^{a_l} \preceq b. \tag{5}$$

By lemma 1 we know that

$$v_{\sigma_a^{(n)}}(b) = 1 \iff a \preceq b. \tag{6}$$

On the other hand, let $h = \sigma_{2^{a_1}}^{(n)} \cdots \sigma_{2^{a_l}}^{(n)}$, then

$$v_h(b) = 1 \iff v_{\sigma_{2^{a_1}}^{(n)}}(b) = 1, \ldots, v_{\sigma_{2^{a_l}}^{(n)}}(b) = 1 \iff 2^{a_1} \preceq b, \ldots, 2^{a_l} \preceq b. \tag{7}$$

By (5), (6) and (7) we obtain that

$$v_{\sigma_a^{(n)}}(b) = 1 \iff v_h(b) = 1$$

That is $\sigma_a^{(n)} = \sigma_{2^{a_1}}^{(n)} \cdots \sigma_{2^{a_l}}^{(n)}$.

It is apparent that $\sigma_a^{(n)} \cdot (\sigma_{2^{a_1}}^{(n)} + 1) = \sigma_{2^{a_1}}^{(n)} \ldots \sigma_{2^{a_l}}^{(n)} \cdot (\sigma_{2^{a_1}}^{(n)} + 1) = 0$. $\deg(\sigma_{2^{a_1}}^{(n)} + 1) = 2^{a_1}$. So, if $\lambda_f(2^{t-1}) = \lambda_f(2^t) = \lambda_f(2^t + 2^{t-1}) = 0$, let

$$g = (\sigma_{2^0}^{(n)} + 1)(\sigma_{2^1}^{(n)} + 1) \cdots (\sigma_{2^{t-2}}^{(n)} + 1),$$

then $f \cdot g = 0$, and $\deg(g) = 2^0 + 2^1 + \ldots + 2^{t-2} = 2^{t-1} - 1 < \dfrac{n}{2} \leq \left\lceil \dfrac{n}{2} \right\rceil$. This contradicts $AI(f) = \left\lceil \dfrac{n}{2} \right\rceil$. □

## 5 Conclusion

In this correspondence, we study the property of algebraic immunity of symmetric Boolean functions. We prove that there is exactly one function with maximum algebraic immunity on odd number of variables among trivial balanced symmetric Boolean function. we also obtain a necessary condition for the SANF of a symmetric Boolean function with maximum algebraic immunity.

Finally, there are some interesting directions for future research such as characterize the symmetric functions on even number of variables with maximum algebraic immunity (The existence of several symmetric functions on even number of variables with maximum algebraic immunity have been proved in [12,13]), how to construct non-symmetric Boolean functions (which are not obtained by applying affine transformation on the input variables of symmetric Boolean functions) with required algebraic immunity keeping in mind the other cryptographic criteria.